\documentclass[fleqn,10pt]{wlscirep}
\usepackage[utf8]{inputenc}
\usepackage[T1]{fontenc}

\newcommand{\vk}{{\mathbf{k}}}
 \newcommand{\vq}{{\mathbf{q}}}

\title{Single-particle excitations in the uniform electron gas by diagrammatic Monte Carlo}

\author[1,*]{Kristjan Haule}
\author[1,2]{Kun Chen}
\affil[1]{Department of Physics and Astronomy, Rutgers University,  Piscataway, New Jersey 08854, USA}
\affil[2]{Center for Computational Quantum Physics, Flatiron Institute, 162 5th Avenue, New York, NY 10010. The Flatiron Institute is a division of the Simons Foundation}

\affil[*]{haule@physics.rutgers.edu}

\begin{abstract}
  We calculate the single-particle excitation spectrum and the Landau Fermi
  liquid parameters for the archetypal model of solids, the three-dimensional uniform electron gas,
  with the numerically exact variational diagrammatic Monte Carlo
  method. In the metallic range of density, we establish benchmark
  values for the wave-function renormalization factor $Z$, the
  effective mass $m^*/m$, and the Landau parameters $F_0^s$ and
  $F_0^a$ with unprecedented accuracy, and we resolve the
  long-standing puzzle of non-monotonic dependence of mass on density.
  We also exclude the possibility that experimentally measured large
  reduction of bandwidth in Na metal can originate from the charge and
  spin fluctuations contained in the model of the uniform electron
  gas.
\end{abstract}
\begin{document}

\flushbottom
\maketitle
%
%
\thispagestyle{empty}

\section*{Introduction}

The uniform electron gas
(UEG) is the most fundamental model for understanding the electronic properties of metallic materials. The ground-state properties of the model have been very precisely calculated by quantum Monte Carlo
methods~\cite{Ceperley_MC}, and this allowed one to build approximate density functionals~\cite{LDA,GGA}, which are at the heart of the
ab-initio approaches in material science and modern theory-driven materials design.  The knowledge of the low energy excitations of the same model remain challenging to evaluate accurately~\cite{simion2008,Takada_claim,FinalStateInteraction,Yumi,paramagnons,Ashcroft1,Kukkonen1},
even though such calculations are important for building more sophisticated density functionals~\cite{beyondGGA,MBJohnson,SCAN}, and
these excitations are directly measured in experiments on simple
metals, such as alkaline materials.
{Some aspects of the excitation
spectra, such as the quasiparticle renormalization amplitude, were recently determined by extention of
the variational Monte Carlo method in Ref.~\cite{Ceperley_PRL}, which turn
out to be in remarkable good agreement with our current results.}

In the metallic regime, the low-energy properties of the electron
liquid are dominated by the long-lived quasiparticles near the Fermi
surface, and their dynamics is described by a handful of the Fermi
liquid parameters. These parameters completely characterize the low
energy excitation spectra of the metallic state. Unfortunately, they
are very challenging to calculate by a first principle approach,
therefore they are usually treated as phenomenological
parameters requiring input from experiments.

Here we develop an extension of the recently introduced variational diagrammatic Monte Carlo (VDMC) method~\cite{KunHaule}, which fills
this void, and allows us to determine the single-particle excitations of UEG with unprecedented accuracy.
In this letter, we calculate the single-particle excitation
spectra, and in particular, we give controlled values of the wave-function renormalization factor $Z$, the quasiparticle effective mass ratio $m^*/m$ and also the Landau Fermi liquid parameters
$F_0^a$ and $F_0^s$. Our computed values are free of systematic error, and their uncertainty is
mainly controlled by the statistical error, and hence
our established value can be used as a precise benchmark for new method development. Moreover, these precise Fermi liquid parameters are also
useful for building more sophisticated density functionals. Finally,
the method we develop here can be used to solve more sophisticated
models, and can also be used in the ab-initio framework on models of
realistic materials, a development which is currently
underway~\cite{haule2020all}.


\section*{Results}

\subsection*{The Feynman expansion algorithm}
The VDMC method~\cite{KunHaule} is a flavor of diagrammatic Monte Carlo method (DMC)~\cite{nikolay1998,prokof2008fermi,van2012feynman,VANHOUCKE201095,kozik2010diagrammatic,DMC_Hubbard,rossi2017, rossi2018}, which samples high-order Feynman diagrams with a Monte Carlo importance sampling. The novelty of VDMC is two-fold: i) it optimizes the starting point of the perturbative expansion in such a
way that the expansion converges very rapidly with
the increasing perturbation order. ii) it efficiently combines an exponentially large number of Feynman diagrams, which mostly cancel among themselves due to alternating fermionic sign so that the groups of diagrams
can be efficiently sampled with the Monte Carlo importance sampling
hence avoiding the explosion of statistical error with perturbative
order.


In Ref.~\citenum{KunHaule} we computed the spin and the charge
response functions of the UEG model with VDMC by evaluating the
Feynman diagrams for the polarization function.  A similar type of
Feynman expansion in terms of non-interacting single-particle Green's
function, and statically screened Coulomb interaction does not
converge rapidly enough to establish a reliable infinite order result,
hence we here develop an alternative approach.

\begin{figure}[bht]
\includegraphics[width=0.5\linewidth]{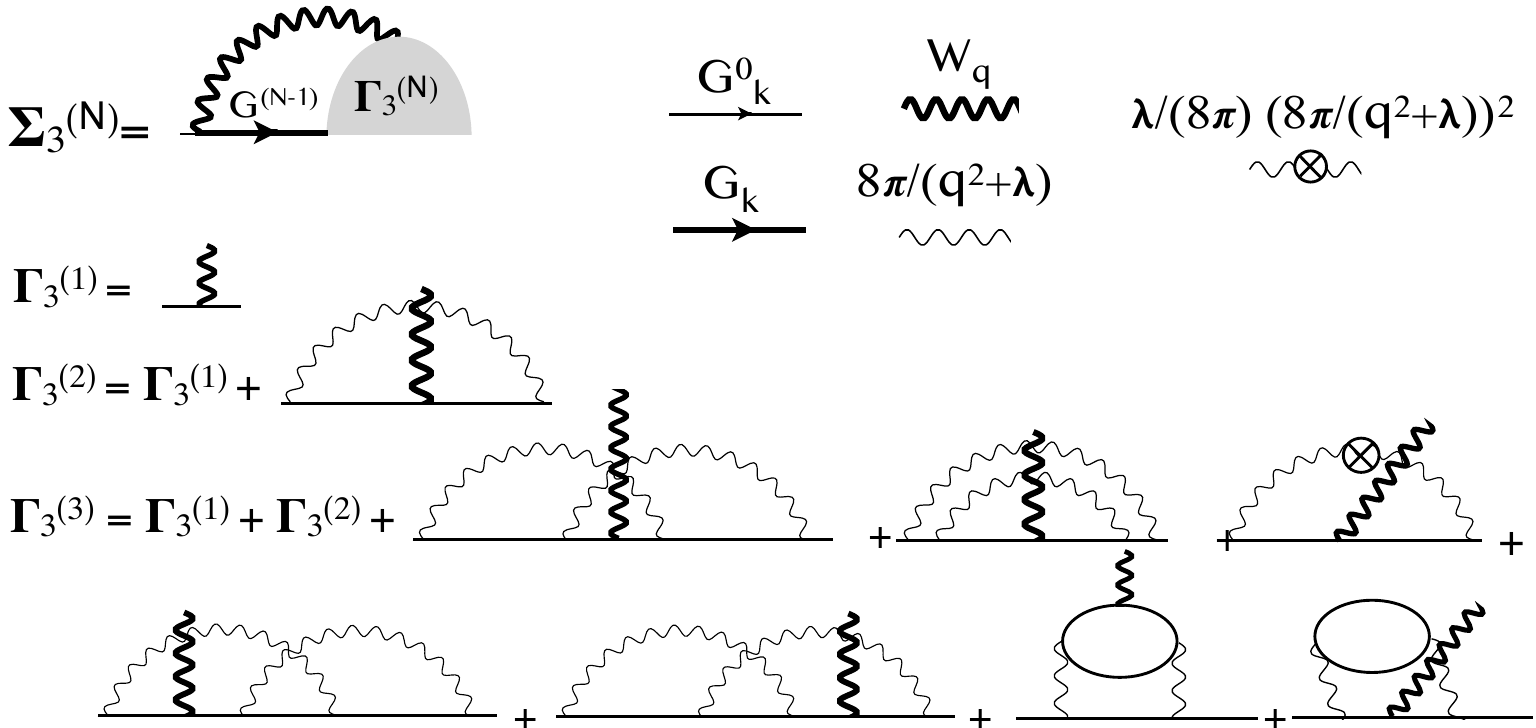}
\caption{
  \textbf{Feynman diagrams} for the self-energy in terms of
  the three leg vertex $\Gamma_3$, which is expanded in bare series in terms of
  $G_\vk^0$ and partially screened interaction
 $v_\vq=\frac{8\pi}{q^2+\lambda}$ and
counter-terms $(\frac{\lambda}{8\pi})^N(\frac{8\pi}{q^2+\lambda})^{N+1}$.
  }
\label{Figm1}
\end{figure}
In this work, we show
that extremely rapid convergence with perturbation order can be
achieved by using a Hedin-type equation, in which we first compute the
numerically exact screened interaction $W_\vq$ (previously developed
in Ref.~\citenum{KunHaule}), and we then expand only the three-point vertex function
$\Gamma_3$ in powers of the bare electron propagator $G^0_\vk$, and 
statically screened interaction $v_q(\lambda)$, with proper counter terms defined
in the Method section. Here the screened Coulomb interaction
$v_q(\lambda)$ has a Yukawa form, characterized by
the inverse screening length $\lambda$.
This screening parameter has to be determined by
the principle of minimal sensitivity in order to achieve rapid
convergence of the perturbative series, so that the extrapolation to
infinite order is possible.
Fig.~\ref{Figm1} shows the sketch of the
corresponding Feynman diagrams up to the third order.
Below we apply the algorithm to the UEG model, although
the method is completely general and could as well be carried out for
realistic material in the ab-initio framework.

\subsection*{The single particle excitations}

\begin{figure*}[bht]
\includegraphics[width=0.5\linewidth]{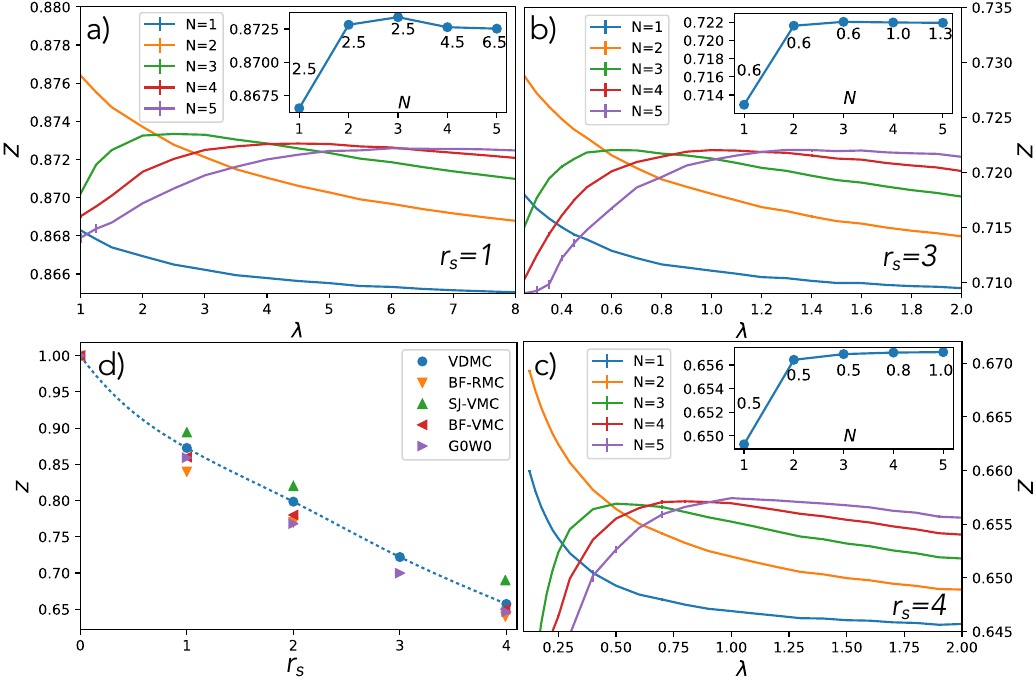}
\caption{ \textbf{The wave-function renormalization factor $Z$} versus
  screening parameter $\lambda$ for various perturbation orders
  $N=1...5$ and for $r_s=1, 2, 3$ and $4$. The insets show the
  convergence of $Z$ with perturbation order $N$ when its value is
  taken at the extremal $\lambda$. The numbers next to each point show the value
  of $\lambda$ used for each calculated point.
  {Panel d) compares current VDMC results with prior Monte Carlo
  results from Ref.~\cite{Ceperley_PRL} and G0W0 from Ref.~\cite{PhysRev.139.A796}.}
}
\label{Fig0}
\end{figure*}
We first present the single-particle excitation spectral results.
Fig.~\ref{Fig0}a-c show how the wave-function renormalization factor $Z$ depends on the screening parameter $\lambda$ in our theory. To determine the optimized parameter $\lambda$, we scan $Z(\lambda)$ for each $r_s$, and determine it with the principle of minimal sensitivity. For efficiency, we here sample the self-energy only at the Fermi wave vector $k_F$ and at the two lowest Matsubara frequencies, which is sufficient to determine $Z$. We notice that for the first two orders, no counter term in the parameter $\lambda$ occurs, therefore the curve $Z(\lambda)$ displayed in Fig.~\ref{Fig0} does not have extremum, while all higher-order terms have a well-defined maximum, which broadens and develops into a broad plateau with increasing order. The insets of Figs.~\ref{Fig0}a-c show optimized $Z$ versus perturbation order, where the first two orders are evaluated at the optimal $\lambda$ of the third order, and for later orders, we take the value in the maximum. We also display the value of $\lambda$ used at each order. From Fig.~\ref{Fig0} it is apparent that beyond order three the rate of convergence to limiting value of $Z$ is extremely fast, and therefore we can confidently determine the first three digits of $Z$. The values and the estimated error-bar from the extrapolation and statistical errors are shown in Table~\ref{tab1}.


\begin{table}
\begin{center}
\begin{tabular}{|c|c|c|c|c|}
  \hline
  $r_s$ &   $Z$ & $m^*/m$ & $F_0^a$ & $F_0^s$\\
  \hline
  1 & 0.8725(2) & 0.955(1) & -0.171(1)& -0.209(5)\\
  \hline
  2 & 0.7984(2) & 0.943(3) & -0.271(2)& -0.39(1)\\
  \hline
  3 & 0.7219(2) & 0.965(3) & -0.329(3)& -0.56(1)\\
  \hline
  4 & 0.6571(2) & 0.996(3) & -0.368(4)& -0.83(2)\\
  \hline
\end{tabular}
\end{center}
\caption{
\textbf{Landau liquid parameters:} The wave-function renormalization factor
$Z$, effective mass $m^*/m$, and the Landau parameters $F_0^a$,
$F_0^s$ for various values of the density parameter $r_s$ with the
estimated error.
}
\label{tab1}  
\end{table}
{In Fig.~\ref{Fig0}d we compare our computed $Z(r_s)$ with the previous
best available estimates, obtained by various flavors of Monte Carlo
(MC) methods, which are reproduced from Ref.~\cite{Ceperley_PRL}.
Note that all these published MC methods rely on fixed
node approximation and the thermodynamic limit extrapolation, hence
they have an inherent systematic error, nevertheless they turn out to
be in very good agreement with current VDMC results. } Our current work based on VDMC
has only statistical error, and a small error in extrapolating in
perturbation order, and is thus far more precise than previous best
results. We notice that previous MC results are broadly consistent
with our results, with SJ-VMC method predicting slightly too large and
BF-VMC and BF-RMC slightly too small value. It is also well known that
G0W0 predicts quite accurate $Z$ values, however, we can now
confidently claim that in the range of metallic densities, G0W0
consistently underestimates $Z$.


\begin{figure}[bht]
\includegraphics[width=0.5\linewidth]{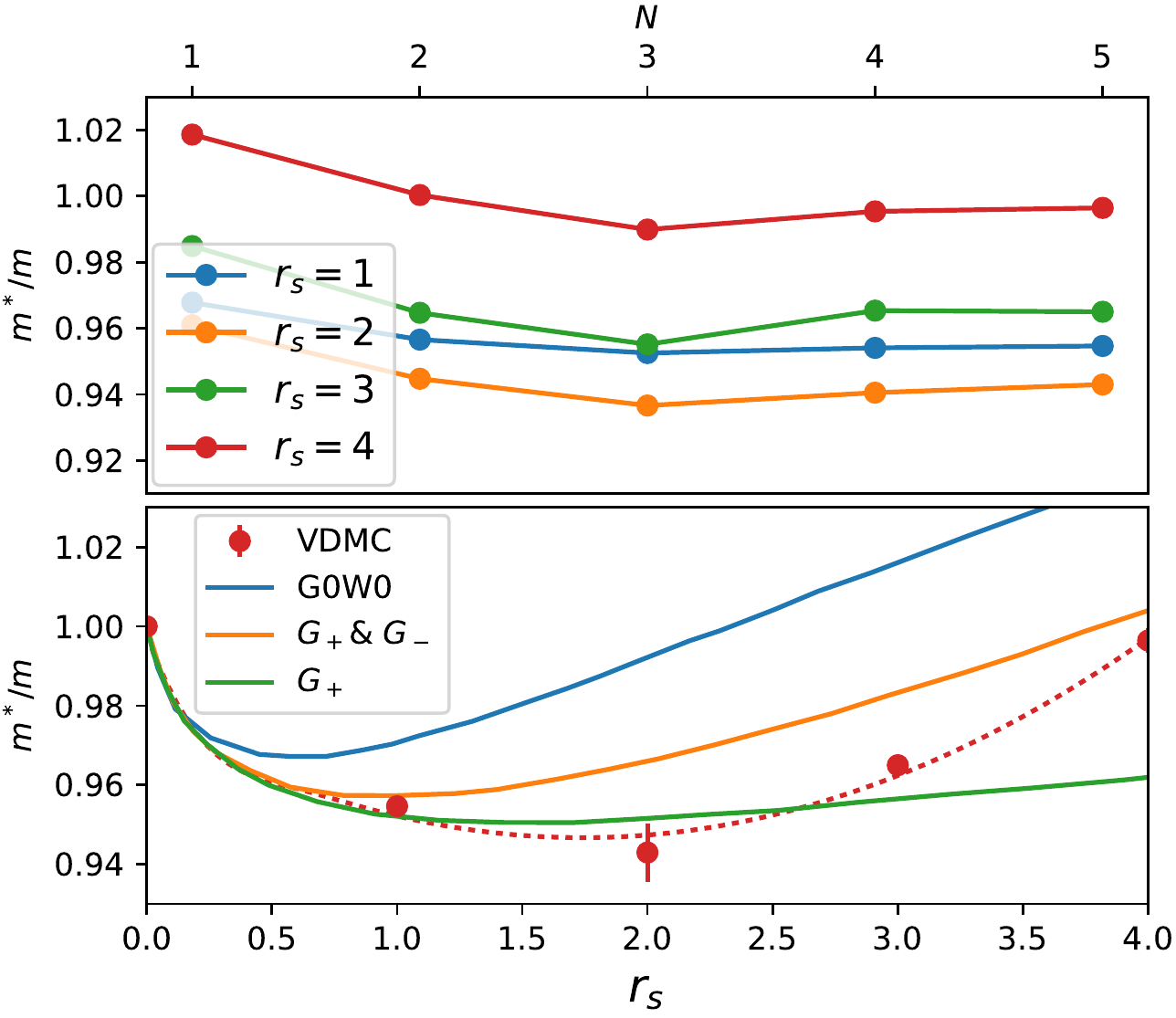}
\caption{
\textbf{Electron effective mass:} The upper panel shows our calculated
effective mass versus perturbation order for $r_s=1-4$. The lower
panel compares the $r_s$ dependence of the effective mass of this work (VDMC)
with the prior analytic and numeric work from Ref.~\citenum{LFT1}.
}
\label{Fig2}
\end{figure}
Once the extremal value of $\lambda$ is determined, we compute the
entire momentum and frequency dependence of the self-energy, which allows
us to determine also the momentum derivative of the self-energy, and
hence the effective mass of the electron through the relation
\begin{eqnarray}
 \frac{m}{m^*} = Z \left(1 + \frac{m}{k_F}\frac{d\Sigma(k_F,\omega=0)}{dk}\right)
\end{eqnarray}
The convergence of the effective mass ratio $m^*/m$ with perturbation order
is shown in Fig.~\ref{Fig2}a, and its dependence on $r_s$ is displayed
in Fig.~\ref{Fig2}b.

The dependence of the effective mass $m^*/m$ on $r_s$ has been
controversial for many decades. Some theories predict that the ratio
is monotonically decreasing with increasing
$r_s$~\cite{Yumi,Lam}, while others predict the existence of a
turning point $r_s^*$~\cite{LFT1,Hubbard,Rice, Hedin_Ludqvist,MacDonald_1980} at which the trend is
reversed. Our controlled results confirm the correctness of the later
theories. Furthermore, we compare our controlled VDMC results with
previous best estimates, which are based on the theory of many-body
local field factors~\cite{LFT1}. This theory includes vertex
corrections associated with charge and spin fluctuations, extracted
from available Monte Carlo data. We notice that G0W0 overestimates the
effective mass in the entire range of metallic densities. The density
fluctuations beyond RPA are included in theory with $G_+$ local field
corrections, which reduce the mass substantially and bring it very
close to our VDMC results at small $r_s$. However, beyond $r_s>3$ our
VDMC results are closer to the theory which contains both the charge
and the spin fluctuations ($G_+ \& G_-$), hence we can infer that at
moderate correlations strength, the spin fluctuations start to play an
important role, and charge fluctuations are no longer sufficient in
determining the mass of the electron gas.


\subsection*{The Landau liquid parameters}
With precisely calculated effective mass, as well as the spin and
charge susceptibility determined in our previous work~\cite{KunHaule},
we can calculate Landau parameters $F_0^a$ and $F_0^s$, which are
obtained from $\frac{\chi_s}{\chi_s^0} = \frac{m^*}{m}\frac{1}{1+F_0^a}$
and $\frac{P_{q=0}}{P_{q=0}^0} = \frac{m^*}{m}\frac{1}{1+F_0^s}$ .
Here $\chi_s$ and $P_q$ are the spin susceptibility and charge
polarization, while $\chi_s^0$ and $P_q^0$ are their non-interacting analogues.
In table~\ref{tab1} we list our calculated Landau parameters $F^a_0$
and $F^s_0$, together with the estimation of their error, which mostly
comes from error in determining spin and charge susceptibility in
Ref.~\citenum{KunHaule}.
While the Landau parameters,
which determine the interaction between
quasiparticle, have been
estimated by various approximate numerical methods
before~\cite{Yumi}, to our knowledge their numerically controlled value has not be
obtained before.

\subsection*{The spectral function and the bandwidth}
\begin{figure}[bht]
\includegraphics[width=0.5\linewidth]{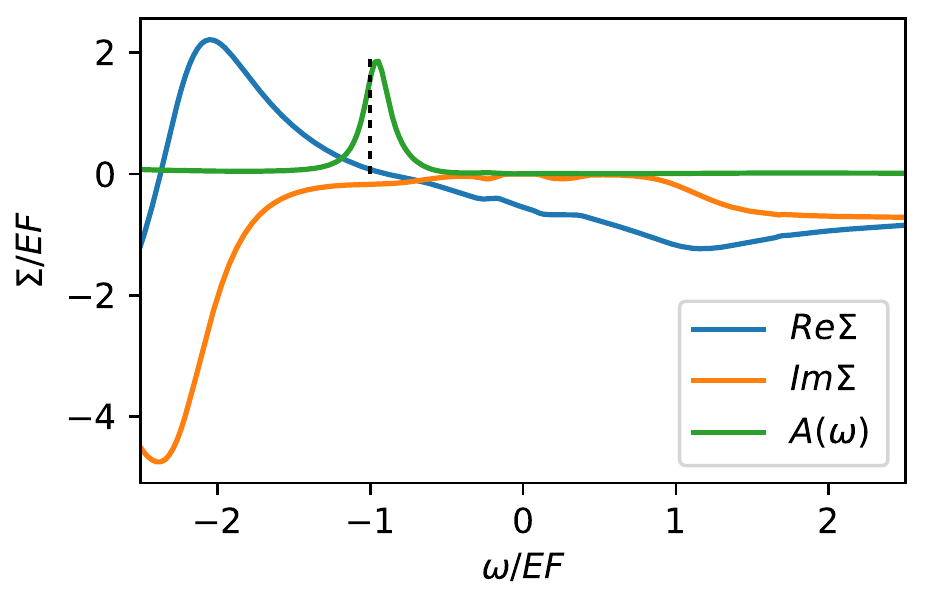}
\caption{
  \textbf{The spectral function} and $\Sigma_{k=0}(\omega)$ at $r_s=4$ and $k=0$ as relevant for
  bandwidth of Na metal.
%
}
\label{Fig3}
\end{figure}
The present VDMC algorithm also allows us to compute a numerically controlled value for the dynamic self-energy on the imaginary axis. Analytic continuation is needed to obtain the self-energy on the real frequency axis. We use the maximum entropy method to compute the quasiparticle energy at the $k=0$ point, which determines the bandwidth of the electron dispersion, i.e., the energy difference between the Fermi level and the lowest possible quasiparticle energy.  In Fig.~\ref{Fig3} we display the self-energy, as well as the spectral function at momentum $k=0$ and finite frequency. We notice that the imaginary part of the self-energy starts to grow rapidly when the energy of the single-particle excitations exceeds the plasma frequency $\omega_p\approx 1.881 E_F$. Consequently, there appears a strong pole at $\omega \approx -2.4 E_F$ due to such plasma excitations, and makes quasiparticle approximation invalid at a frequency below $\omega < -E_F$, as the real part of the self-energy is no longer a linear function of frequency. However, around $E_F$ the real-part of $\Sigma$ is still quite close to a linear function, and only minor deviations are noticed. Consequently, the renormalization of the dispersion can not substantially deviate from our earlier estimation of $m^*/m$, which is valid at the Fermi level. Our numerical estimation based on the analytically continued self-energy is that the spectral function at $r_s=4$ and $k=0$ has a peak around $-0.96 E_F$, which deviates from the non-interacting value for only 4\%, hence the bandwidth reduction due to interactions at $r_s=4$ is only of the order of 4\%. This value is much smaller than the experimental estimation of the bandwidth reduction in Na metal, in which the measured ARPES bandwidth appears to be renormalized for about 18-25\%~\cite{Na_ARPES,Na_ARPES2}. However, our estimated bandwidth is definitely not substantially larger as compared to the non-interacting bandwidth, in contrast to several other many-body calculations~\cite{QMC_Na,Takada}, and is neither substantially smaller as in early GW calculations~\cite{Hybertsen} or GW with paramagnon vertex corrections~\cite{paramagnons}. Based on our very precise estimation of the single-particle self-energy, we can confidently exclude a possibility of such a dramatic reduction of the bandwidth in the model of electron gas due to correlation effects at the density corresponding to Na metal. This large reduction of the effective mass in ARPES thus requires an alternative explanation, which was assigned to the interaction in the final states~\cite{Mahan_Na,Takada} in ARPES, surface effects~\cite{Mahan_Na2}, and possibly the lattice effects, i.e, deviation of Na metal from the continuous model of the uniform electron gas. 

In summary, we established the low energy excitation spectrum of the uniform
electron gas at metallic density using recently developed
VDMC. Controlled values of $Z$, $m^*/m$, $F_0^s$, and $F_0^a$ are
given, which agree with the state of the art calculations in the field,
but here we provide much more precise values than previously known.


\section*{Methods}
The Hamiltonian of UEG problem is
\begin{eqnarray}
\hat{H}=\sum_{\vk\sigma} \left( {\vk^2}-\mu \right) \hat{\psi}^\dagger_{\vk\sigma}\hat{\psi}_{\vk\sigma}+
\nonumber\\
\frac{1}{2V}\sum_{\substack{\vq\ne 0\\ \vk\vk'\sigma\sigma'}}
\frac{8\pi}{q^2}\hat{\psi}^\dagger_{\vk+\vq\sigma}\hat{\psi}^\dagger_{\vk'-\vq\sigma'}\hat{\psi}_{\vk'\sigma'}\hat{\psi}_{\vk\sigma},
\label{H_UEG}
\end{eqnarray}
where $\hat{\psi}$/$\hat{\psi}^\dagger$ are the annihilation/creation
operator of an electron, $\mu$ is the chemical potential controlling
the density of the electrons in the system, and the long-range Coulomb
repulsion is $8\pi/q^2$, as we measure the energy in units of
Rydbergs, and the wave number $k,q$ in units of inverse Bohr radius.

The expansion in terms of the bare interaction is divergent, therefore
we first transform the original problem into an equivalent but a
more appropriate problem for power expansion, which describes the
emergent degrees of freedom at the lowest order, and the corrections
are perturbatively included with very rapid convergence.
Motivated by the well-known fact that the long-range Coulomb
interaction is screened in the solid and that the effective potential
of emerging quasiparticles differs from the bare potential, we
introduce the screening parameter $\lambda_\vq$ and an electron
potential $v_\vk$ into the quadratic part of the emergent Lagrangian
$L_0$ of the form
\begin{eqnarray}
L_0 = \sum_{\vk\sigma}\psi^\dagger_{\vk\sigma}
\left(\frac{\partial}{\partial\tau}-\mu+\vk^2+v_\vk(\xi=1)\right)\psi_{\vk\sigma}
\nonumber\\
+\sum_{\vq\ne 0} \phi_{-\vq}
  \frac{q^2+\lambda_\vq}{8\pi}\phi_\vq, .
\label{L0}
\end{eqnarray}
We then add the following interacting part to the Lagrangian
\begin{eqnarray}
\Delta L = 
-\sum_{\vk\sigma}\psi^\dagger_{\vk\sigma} v_\vk(\xi)\psi_{\vk\sigma}
-\xi \sum_{\vq\ne 0}\phi_{-\vq}
  \frac{\lambda_\vq}{8\pi}\phi_\vq
\nonumber\\
+\sqrt{\xi}\frac{i}{\sqrt{2 V}}\sum_{\vq\ne 0} \left(\phi_{\vq}\rho_{-\vq}+\rho_{\vq}\phi_{-\vq}\right).
\label{deltaL}
\end{eqnarray}
so that, when the number $\xi$ is set to unity, $L(\xi) = L_0(\xi)+\Delta L(\xi)$ is Lagrangian of UEG. Indeed integrating out the bosonic
fields $\phi_\vq$ from Lagrangian $L$, we get the Lagrangian corresponding to the original Hamiltonian Eq.~(\ref{H_UEG}). Here $\rho_\vq$ is the density fluctuation of the problem $\rho_\vq=\sum_{\vk\sigma}\psi^\dagger_{\vk\sigma}\psi_{\vk+q\sigma}$. Note that the first two terms in $\Delta L$ are the counterterms~\cite{Counterterm} 
which exactly cancel the two terms we added to $L_0$ above.  We use the number $\xi$ to track the order of the Feynman diagrams so that order $N$ contribution sums up all diagrams carrying the factor $\xi^N$. We set $\xi=1$ once we enumerate all the diagrams of a certain order.

The emergent screening length $\lambda_\vq$ and effective potential $v_\vk$ are not a-priory known and need to be properly optimized to achieve an optimal speed of convergence. We note in passing that
determining those parameters self-consistently, i.e., $\lambda_\vq$ from the self-consistent polarization, and $v_\vk$ from the single-particle self-energy, is not the most optimal choice for the speed of convergence. Determining them by the principal of minimal sensitivity is a much better choice, as pointed out by Kleinert and Feynman~\cite{stevenson1984,stevenson1985,stevenson1986,kleinert1995}. They showed that when an effective parameter of a theory is optimized with this principle, the perturbative expansion converges very fast, and can succeed even when the interaction is strong, and regular perturbation theory fails.

To make algorithm sufficiently simple to implement, we take
$\lambda_\vq$ to be $\vq$ independent constant ($\lambda$), which is
already sufficient for rapid convergence of the series. We emphasize
that for any choice of these parameters we are guaranteed to converge
to the same answer, provided that the series converges. Furthermore,
we found that the convergence of the expansion is best when the Fermi
surface of both the dressed $G_\vk$ and the bare $G^0_\vk$ Green's
function at each order is fixed with the Luttinger's theorem so that
the density and the Fermi surface volume is not changed with the
increasing perturbation order. We therefore, expand $v_\vk$ in power
series
$v_\vk= (\Sigma^x_\vk(\lambda)-\Sigma^x_{k_F}(\lambda)) + \xi\, s_1 +
\xi^2\, s_2\cdots$, and we determine $s_N$ so that all contributions
at the order $\xi^N$ do not alter the physical volume of the Fermi
surface. Similarly to optimizing $\lambda_\vq$, we found that it is
sufficient to take $s_N$ constants independent of the momentum. Since
the exchange ($\Sigma^x_\vk$) is static and is typically large, we
accommodate it at the zeroth-order into the effective potential, so
that at the first order we recover the GW type self-energy with
$G_\vk$ at the screened Hartree-Fock (screened by screening length
$\lambda$) and exact $W_\vq$.

As mentioned before, the algorithm depicted in Fig.~\ref{Figm1}  needs
a numerically exact (converged) $W_\vq$, which is first computed with
the algorithm of Ref.~\citenum{KunHaule}. It was shown in
Ref.~\citenum{KunHaule} that the most rapidly converging scheme for
charge and spin-susceptibilities is the so-called vertex correction
scheme, in which we add an infinite sum of ladder diagrams on both
sides of a polarization Feynman diagram. To do that, we first
precompute the three-point ladder vertex and then attach it to both
sides of a polarization Feynman diagram while the diagrams are
sampled, and at the same time, we eliminate all ladder-type diagrams
from the sampling, to avoid double-counting of diagrams. Next, we use
Hedin's type equation depicted in Fig.~\ref{Figm1} in which one
fermion propagator is dressed and requires self-consistent $G$. It
easy to see that it is sufficient to use bold $G$ of the lower order
$N-1$ when evaluating self-energy at order $N$, to avoid the expensive
self-consistent calculation. Finally, we use the finite temperature
imaginary-time formalism, and we set the temperature to  $T=0.04
\,E_F$, which is sufficiently below the Fermi liquid scale, so that is
essentially equivalent to zero temperature.

\bibliography{vDiagMC}{} 



\section*{Acknowledgements}

This work was supported by the Simons Foundation through Simons fellowship and Simons Collaboration on the Many Electron Problem. KH acknowledges
supported of NSF DMR-1709229. The Flatiron Institute is a division of the Simons Foundation.

\section*{Author contributions statement}

K.C. and K.H. both developed two independent
codes to crosschecked the results, K.H. wrote the first draft of the
manuscript and K.C. and K.H. contributed to the manuscript and
approved the text.

\section*{Additional information}

\textbf{Accession codes:} The source code has been made available for download under gnu license
at: \url{https://github.com/haulek/VDMC}

\noindent
\textbf{Competing interests:} The authors declare no competing interests.

%
%

\end{document}